\documentclass[a4paper,11pt]{article}
\usepackage{pos}

\let\l\left
\let\r\right

\let\mrm\mathrm

\let\mbb\mathbb

\let\dag\dagger
\newcommand{\beq}{\begin{equation}}
\newcommand{\eeq}{\end{equation}}
\newcommand{\beqa}{\begin{eqnarray}}
\newcommand{\eeqa}{\end{eqnarray}}
\newcommand{\ba}{\begin{array}}
\newcommand{\ea}{\end{array}}
\newcommand{\bmat}{\begin{pmatrix}}
\newcommand{\emat}{\end{pmatrix}}
\newcommand{\bcas}{\begin{cases}}
\newcommand{\ecas}{\end{cases}}
\newcommand{\muh}{{\hat{\mu}}}

\title{Thimble regularisation of YM fields: crunching a hard problem}
\ShortTitle{Thimble regularisation of YM fields}

\author*{Francesco Di Renzo}
\author{Simran Singh}
\author{Kevin Zambello}

\affiliation[]{Dipartimento di Scienze Matematiche, Fisiche e
  Informatiche, Università di Parma\\ 
and INFN, Gruppo Collegato di Parma, I-43100, Parma, Italy}

\emailAdd{francesco.direnzo@unipr.it}
\emailAdd{simran.singh@unipr.it}
\emailAdd{kevin.zambello@studenti.unipr.it}

\abstract{Thimble regularisation of Yang Mills theories is still to a
  very large extent terra incognita. We discuss a couple of topics
  related to this big issue. 2d YM theories are in principle good
  candidates as a working ground. An analytic solution is known, for
  which one can switch from a solution in terms of a sum over
  characters to a form which is a sum over critical points. We would 
  be interested in an explicit realisation of this mechanism in the 
  lattice regularisation, which is actually quite hard to work out. 
  A second topic is the inclusion of a topological term in the lattice 
  theory, which is the prototype of a genuine sign problem for pure YM 
  fields. For both these challenging problems we do not have final
  answers. We present the current status of our study.}

\FullConference{%
 The 38th International Symposium on Lattice Field Theory, LATTICE2021
  26th-30th July, 2021
  Zoom/Gather@Massachusetts Institute of Technology
}

\begin{document}
\maketitle

\section{A thimble primer}
\label{sec:Basics}

QCD at finite baryon density is still to a large extent {\em terra
incognita}, due to the infamous sign problem. The latter is in fact
more general (and fundamental, in a sense): we have to tackle it every
time the action of a quantum field theory is complex valued. A number 
of possible solutions have been put forward, among which thimble 
regularisation \cite{Aurora,Kikukawa}. In a light notation in which 
a field theory looks like an ordinary integral, the thimble approach
to field theories is quite easy to describe in terms of the following
recipes:
\begin{enumerate}
\item We first need to complexify the degrees of freedom, {\em i.e.} $\,x
  \rightarrow z=x+iy$ and $\,S(x) = S^R(x) + iS^I(x)$\footnote{It is
    important to remind that the action is complex since the very
    beginning, even for real degrees of freedom.} $\rightarrow S(z)$.
\item We then need to find the critical points, {\em i.e.} those
  points $p_{\sigma}$ where $\partial_z S = 0$.
\item We define the thimble $\mathcal{J}_\sigma$ attached to each critical point as the union of all
  the Steepest Ascent paths (SA), the latter being the solutions of $\frac{d}{dt} z_i = \frac{\partial
  \bar{S}}{\partial \bar{z_i}}$ stemming from the critical point. 
\item If the Hessian of the action has no zero eigenvalue, one can
  immediately prove that the thimble
  is a manifold of the same real dimension as the original manifold we
  started from.
\item Due to the holomorphic nature of $S$, $S^R$ is increasing along
  the ascent and thus on the thimble the original
  integral is convergent, while $S^I$ stays constant.
\item Sadly, the sign problem is {\em not completely} killed,
since the integration measure (encoding the orientation of the
thimble with respect to the embedding manifold) reintroduces a 
{\em residual} sign problem due the so-called {\em residual phase}.
\end{enumerate}
The last point would deserve much more attention than we can pay here;
the interested reader can look at \cite{thimbleRMT} for our (basic,
but stemming from first principles) solution to the computation of 
the residual phase. 
While going through all the steps needed to carry out a computation on
thimbles can be a non trivial task, our efforts are fully rewarded by 
Lefschetz/Picard theory, which states that a thimble decomposition 
for the original path integral holds
\begin{equation}
<O> \, = \, \frac{\sum_{\sigma} \; n_{\sigma} \,
  e^{-i\,S^I(p_{\sigma})} \, \int_{\mathcal{J}_\sigma} dz \;
  e^{-S^R}\; O\; e^{i\,\omega}}{\sum_{\sigma} \; n_{\sigma} \, e^{-i\,S^I(p_{\sigma})}\, \int_{\mathcal{J}_\sigma} dz \;
  e^{-S^R}\; e^{i\,\omega}}
\label{eq:ThimbleDecomposition}
\end{equation}
In (\ref{eq:ThimbleDecomposition}) $S^I$ is no longer such a big problem, 
while the residual sign problem is due to the residual phases
$e^{i\,\omega}$. Notice that both the numerator
and the denominator ({\em i.e.} the partition function) receive
contributions in principle by all the critical points. This is not
really the case, since the {\em
intersection numbers} $n_{\sigma}$ can be zero for possibly many
critical points. It can be shown that $n_{\sigma}=0$ for a critical point when the
associated {\em unstable} thimble does not intersect the original
integration manifold\footnote{The unstable thimble is defined as the union of
the Steepest Descent (SD) paths stemming from a critical point.}.\\

\section{Thimble regularisation of gauge theories}

We started our discussion on motivations for thimbles putting forward
the big issue of QCD at finite density. How far are we from actually
tackling that? Honestly, quite a lot. We have in recent years taken
some steps in that direction, but that has been done in the context of
two theories ($0+1$ QCD \cite{QCD01} and the so called {\em Heavy Dense
  QCD} \cite{Zambello:2018ibq}) for which gauge invariance in
practice does not show up in its full glory. Other groups have perhaps
moved a bit further than we have done till now 
\cite{Alexandru:2018ngw,Pawlowski:2021bbu}. In the following we will
try to pin down a sort of status report on our attempts at a thimble
regularisation of gauge theories. This will make us discuss $2D$ YM
theories and the inclusion of a $\theta$-term.

\subsection{Construction of the thimble}
\begin{figure}[ht] 
\centering
\includegraphics[scale=0.45]{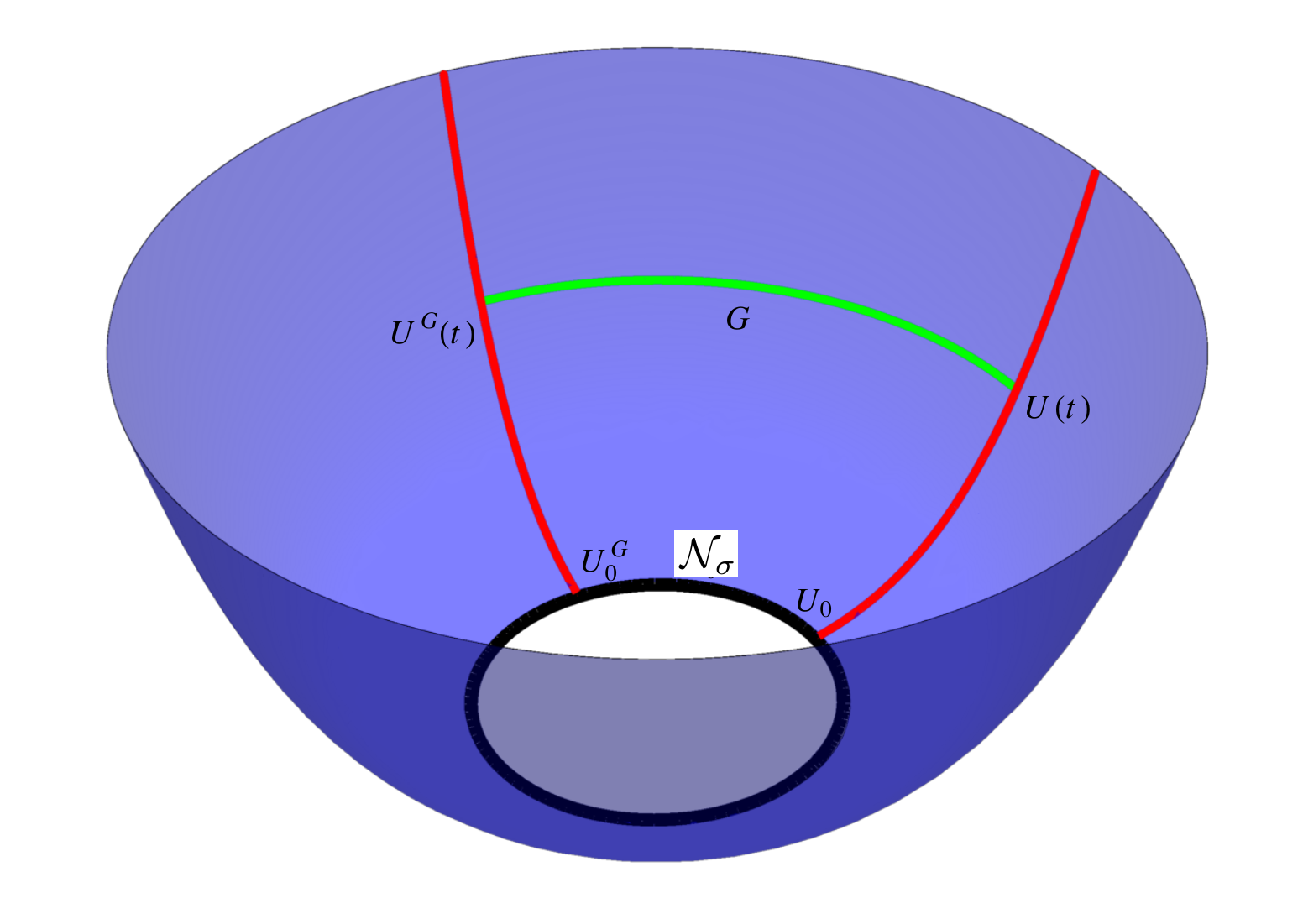}
\caption{The cartoon for the gauge invariance property of the
  construction of Eq.~{\ref{eq:YMthimble}}.}
\label{fig:scodella}
\end{figure}

Mimicking thimble construction for gauge theories is not that
difficult. The first step (complexification) amounts to
\begin{equation}
\mrm{SU}\l(N\r)\ni U=e^{i x_aT^a}\rightarrow e^{i z_aT^a}=e^{i \l(x_a+i y_a\r)T^a}\in\mrm{SL}\l(N,\mbb{C}\r).
\end{equation}
The main thing we should notice is that
$$
\mathrm{SU}\left(N\right)\ni U^\dag=e^{-i x_aT^a}\rightarrow e^{-i z_aT^a}=e^{-i \left(x_a+i y_a\right)T^a}=U^{-1}\in\mathrm{SL}\left(N,\mathbb{C}\right).
$$
With this caveat in mind, we can proceed to defining the SA
\begin{equation}\label{eq:SAgauge}
\frac{\mrm{d}}{\mrm{d}\tau}U_\muh\l(n;\tau\r)=\l(i\,T^a\bar{\nabla}_{n,\muh}^a\overline{S\l[U\l(\tau\r)\r]}\r)U_\muh\l(n;\tau\r)
\end{equation}
which are written in terms of the Lie derivative
$$
\nabla^af\left(U\right)=\lim_{\alpha\rightarrow 0}\frac{1}{\alpha}\left[f\left(e^{i\alpha T^a}U\right)-f\left(U\right)\right]=\frac{\delta}{\delta\alpha}f\left(e^{i\alpha T^a}U\right)\biggl|_{\alpha=0}.
$$
Notice that, since $\frac{\mrm{d}}{\mrm{d}\tau}=
\bar{\nabla}_{n,\muh}^a\bar{S}\,\nabla_{n,\muh}^a+
\nabla_{n,\muh}^aS\,\bar{\nabla}_{n,\muh}^a$ we have that 
$$
\frac{\mrm{d}S^R}{\mrm{d}\tau}=\frac{1}{2}\frac{\mrm{d}}{\mrm{d}\tau}\l(S+\bar{S}\r)=\frac{1}{2}\l(\bar{\nabla}_{n,\muh}^a\bar{S}\,\nabla_{n,\muh}^a S+\nabla_{n,\muh}^aS\,\bar{\nabla}_{n,\muh}^a\bar{S}\r)=\left\Vert\nabla S\right\Vert^2\geq0
$$
and
$$
\frac{\mrm{d}S^I}{\mrm{d}\tau}=\frac{1}{2i}\frac{\mrm{d}}{\mrm{d}\tau}\l(S-\bar{S}\r)=\frac{1}{2i}\l(\bar{\nabla}_{n,\muh}^a\bar{S}\,\nabla_{n,\muh}^a S-\nabla_{n,\muh}^aS\,\bar{\nabla}_{n,\muh}^a\bar{S}\r)=0,
$$
that is, the main properties we expect from the SA are satisfied. 
Among all the solutions of Eq.~(\ref{eq:SAgauge}), we have to
look for the ones whose union defines the thimble. Naively, we could
think we need to
consider those stemming from a critical point $U_\sigma$. The fact is critical
points in gauge theories come along with an entire orbit, which is
made by all the gauge replicas of the given critical point. This
is not the end of the story, since complexification has left us with
two possible candidates. The action is now invariant under the gauge
group ${\cal G}=SL(N,\mbb{C})$, so that one orbit we could think of is
$$
{\cal M_\sigma} = \{U \in SL(N,\mbb{C}) \,|\, \exists \,G\in SL(N,\mbb{C}) :
  U_\sigma^G = U\}.
$$  
This is not the right choice, and we need to consider instead
\beq
{\cal N_\sigma} = \{U \in SL(N,\mbb{C}) \,|\, \exists \,G\in SU(N) :
  U_\sigma^G = U\}\; \subset \;{\cal M_\sigma}.
\eeq
In general the relevant gauge group is ${\cal H}=SU(N)$, and as a
matter of fact the thimble itself is invariant under $SU(N)$ (as it
should be), and not under $SL(N,\mbb{C})$. 
All in all, the thimble {\em e.g.} associated to $A=0$ for the
$SU(3)$ Yang-Mills action is defined by\footnote{We denote ${\cal
    N}^{(0)}$ the orbit ${\cal N_\sigma}$ associated to $A=0$.}
\beq \label{eq:YMthimble}
{\cal J}_0:= 
\left\{ 
U\in (SL(3,\mathbb{C}))^{4V} \; | \; \; 
\exists U(\tau) \; \; \mbox{ solution of Eq.~(\ref{eq:SAgauge})}
\; \; | \; \; 
U(0)=U 
\; \; \& \; \; 
\lim_{\tau\rightarrow - \infty} U(\tau) \in {\cal N}^{(0)} 
\right\}.
\eeq
One could think that what we have gone through till now can be summarised as ``going from
critical points to critical submanifold''. We have to
admit we have been cheating a little bit: what we actually did 
(and this is actually the right thing to do) was to
go from {\em non-degenerate} critical points to {\em non-degenerate} 
critical submanifolds \cite{AtBott}. In Sec.~\ref{sec:Basics} we said
that in order for the thimble
construction to work we need a non-degenerate critical point, with no
zero eigenvalue in the Hessian. Due to gauge modes, this
is not the case for gauge theories. The good news is that this is not such a big problem:
gauge invariance is realised in a very neat way on the thimble.
Indeed the main gauge invariant property of the construction is summarised
in the cartoon\footnote{We provide a
  quick-and-dirty argument for what a non-degenerate critical
  submanifold is; all this can be much better
  understood from \cite{AtBott}.} of Fig~\ref{fig:scodella}.
One can spot the point $U(t)$ which is obtained by ascending from the
critical point $U_0$\footnote{We are still thinking of the critical
  point $A=0$, which in the Wilson action is associated to its
  exponential $U_0$.}: this point is uniquely defined by selecting a given
direction on the tangent space to the thimble at $U_0$\footnote{This
  is always known by solving a convenient Takagi problem; see
  \cite{thimbleRMT}.} and a given ascend time $t$. Ascent paths of
this type are defined selecting directions associated to positive
eigenvalues of the Hessian. As for (zero) gauge modes, their role is
well understood by looking at the point $U^{G}(t)$. All in all, 
if we take a SA from $U_0$ ({\em i.e.} $A=0$), at any stage ({\em i.e.} from any $U(t)$) we can 
perform a gauge transformation $G$ and this will take us to a point
($U^{G}(t)$) starting from which the SD (Steepest Descent) path will
make us eventually 
land on another point on the gauge orbit attached to $U_0$;
this point ($U_0^{G}$) is obtained from $U_0$ by the gauge
transformation we choose.\\
All that we have said till now is not the end of the story, and we
have been once again cheating a little bit. In order to preserve 
the right number of zero modes, in YM theories we have to kill 
{\em torons}, which thing can be done by going for twisted boundary 
conditions. Because of this, our preferred critical point is the so-called 
{\em twist-eater}. (A good reference for all that has to do with this
is \cite{torons}.) 

\subsection{What we have, what we can do and what we lack}

\begin{figure}[ht] 
\centering
\includegraphics[scale=0.65]{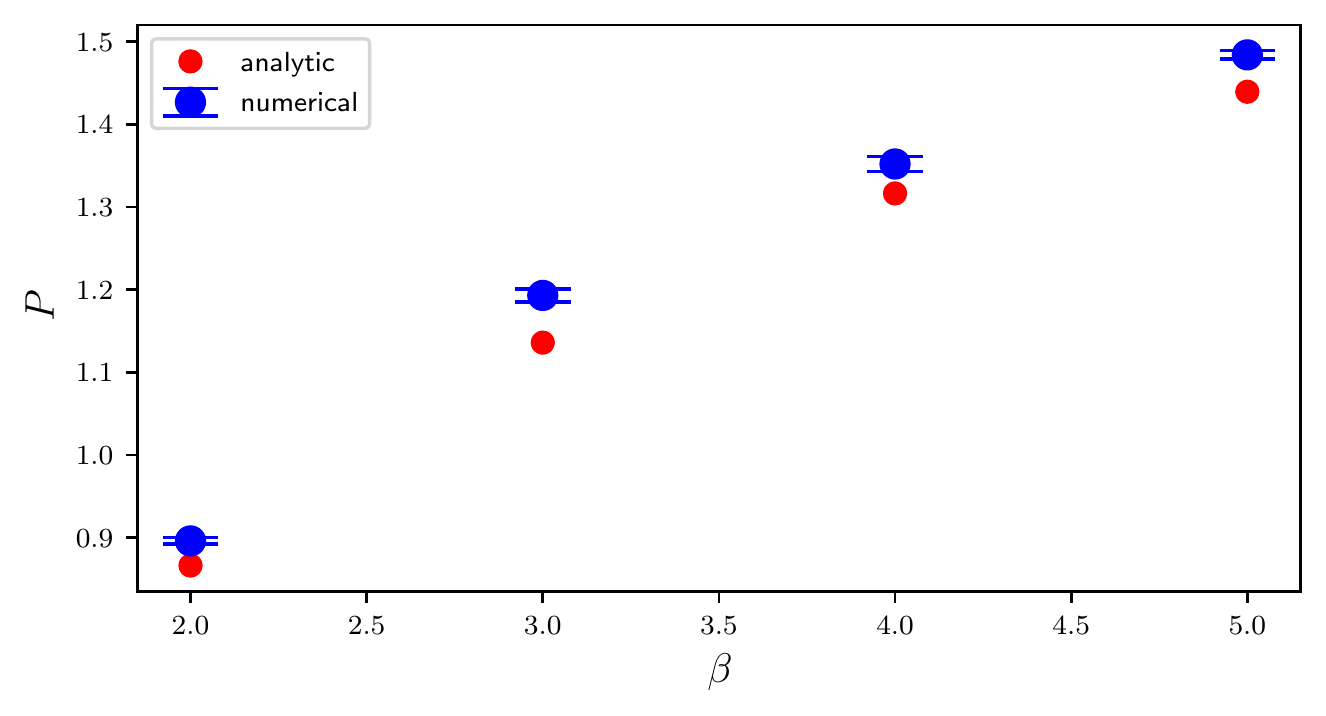}
\caption{Average Plaquette vs real $\beta$ for the numerical (only dominant thimble) vs analytical results for $2D \; SU(2)$ YM theory.}
\label{fig:AveragePlaq}
\end{figure}

We have a (working) code (some results in Fig. \ref{fig:AveragePlaq}) by which we can simulate $2D$ YM theories; to
be definite, we have been mainly focusing on $SU(2)$. A sign problem (an
artificial one, in a sense) is generated by computing for complex
values of the coupling. Notice that the theory is fully solved, so
that we can check the results we get. One thing we do is {\em hunting
  for critical points}, the main goal being to compute on different
thimbles. This is in the logic of (dis)proving whether one single thimble is
enough to reproduce the known result (and this indeed appears not to
be the case: see Fig.~\ref{fig:AveragePlaq}). \\
In all that we have just mentioned there is indeed something we
(dramatically) miss: {\em there is no general proof of something like
  a thimble decomposition for gauge theories}. This is the genuine
motivation of ours for probing $2D$ YM theories: the solution is
known in a way that is intriguing, sort of {\em alluding} to thimbles.

\subsection{The closest to thimble decomposition for gauge theories we can currently
  think of getting}

\begin{figure}[ht] 
\centering
\includegraphics[scale=0.55]{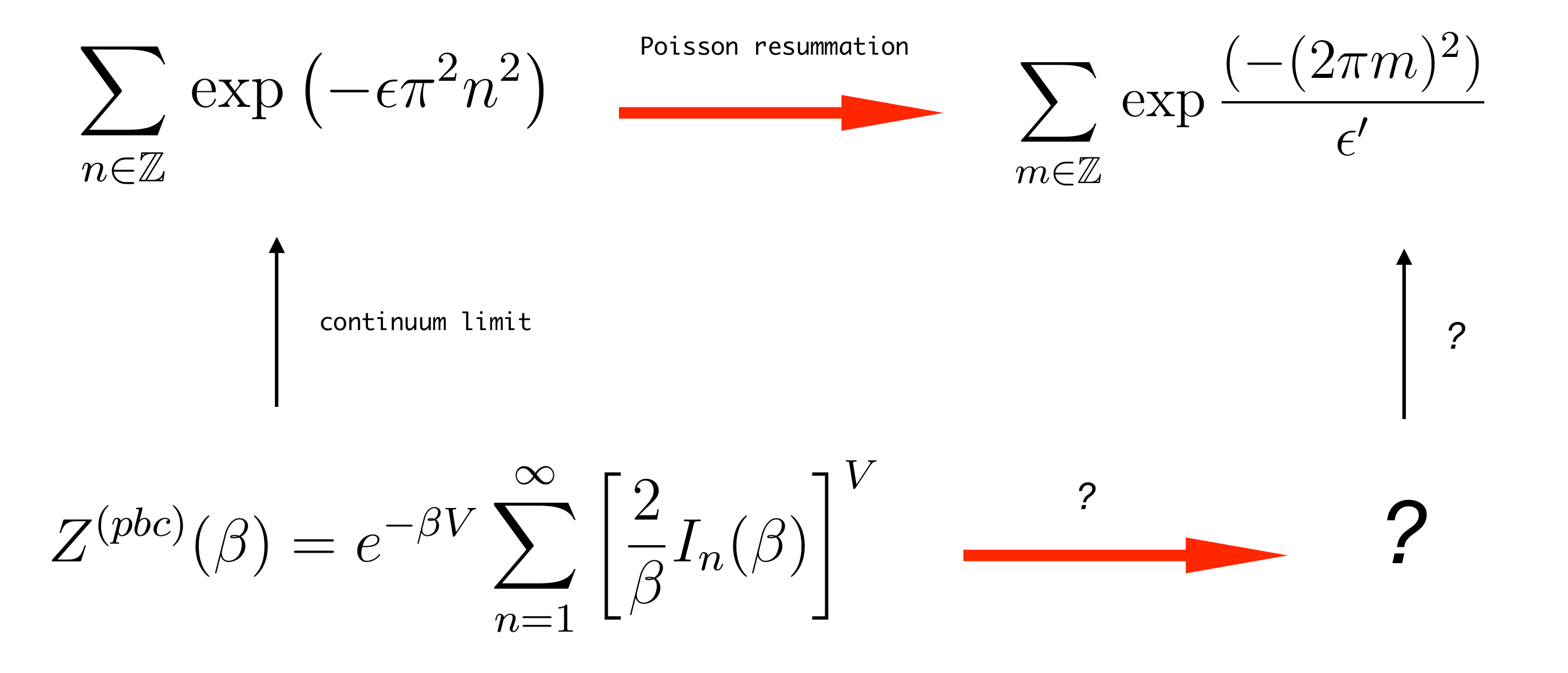}
\caption{The flow chart of a possible argument in favor of thimble
  decomposition in $2D \; SU(2)$ YM theory.}
\label{fig:TheBigIssue}
\end{figure}

Indeed we think $2D$ YM theories provides us with an understanding
which is {\em the closest to thimble decomposition for gauge theories we can currently
  think of getting}. What we mean is displayed (once again) in
graphical form in Fig.~\ref{fig:TheBigIssue}. For a full account of
the results we are going to quote a first reference is \cite{Witten:1992xu}.
First of all look at the first row of Fig.~\ref{fig:TheBigIssue}. All
the critical points that are classical solutions of the YM action have
been classified by Atiyah and Bott in \cite{AtBott}. In
\cite{Witten:1992xu} Witten first obtains the partition function as a sum
over representations (this is a type of result which is well known to
lattice practitioners; see later) on a generic Riemann surface of {\em genus} $g$,
which for $g=1$ reduces to the expression in the up-left corner of 
Fig.~\ref{fig:TheBigIssue}. Via a tool as simple as the {\em Poisson
  resummation}, he then turns this sum into a sum over critical points
(up-right corner). Needless to say, this is the intriguing result we
are mostly interested in: this really {\em sounds like thimble
  decomposition}. 
Now look at the second row. Down-left corner of
Fig.~\ref{fig:TheBigIssue} is the sum over representation which is
well known to the lattice community, {\em i.e.} the Migdal solution
\cite{Migdal:1975zg}. It is well known that one can take the continuum
limit and go from down-left to up-left \cite{Gonzalez-Arroyo:1981ckv}. 
Finally, what we regard as the big issue: can we fill the down-right
corner, {\em i.e.} provide a realisation of the mechanism in the first row in the 
  lattice regularisation? The complete task would entail taking the
  continuum limit and go from down-right to up-right. 

\section{Looking into $\theta$-term}

While we have been working on $2D$ YM theories, we have adapted our
code to also include $4D$ YM in the presence of a $\theta$-term. Once
again we do not have yet definite results, but it is easy to list a
few reasons for being interested in this.
\begin{itemize}
\item This is a prototype of a {\em genuine} sign problem in Euclidean
  Yang Mills.
\item Being the equations of motion unaffected by a $\theta$-term, the
  topological charge is conserved while you ascent on the thimble; in
  a way, this is a genuine way of computing at frozen topological
  charge.
\item We have already made the point that the gauge invariance of the
  thimble is that of $SU(N)$; this holds for the
  topology as well.
\item All in all, the really intruiging (super-hard!) goal would be that of
  finding the weights of the various topological sectors in the
  functional integral.
\end{itemize}
Needless to say, one should be well aware of the effects of lattice
artifacts (they must show up, as in any lattice computations of
topology). And of course, all this is indeed interesting, but it is not at all
guaranteed that we can get positive results in a short time.

\section{Conclusions}
We have provided a status report of our attempts at formulating gauge
theories on thimbles. We do not have positive results to present, but
we have a couple of lines of research that are hard to crunch, but
fascinating. We would like to find a lattice realisation of the
mechanism that in $2D$ gauge theories enables to go from a solution 
in terms of a sum over characters to a form which is a sum over
critical points. $4D$ YM in the presence of a $\theta$-term is a
second subject which is rich of interesting features to investigate.

\section{Acknowledgments}
This work was supported by the European Union Horizon 2020 research and innovation
programme under the Marie Sklodowska-Curie grant agreement No 813942 (EuroPLEx) 
and by the I.N.F.N. under the research project ({\em iniziativa specifica}) QCDLAT.

\end{document}